\documentclass[reprint,superscriptaddress,amsmath,amssymb,aps]{revtex4-1}
\usepackage[utf8]{inputenc}
\usepackage[T1]{fontenc} 
\usepackage{graphicx}
\usepackage{rotating}
\usepackage{makeidx}
\usepackage{threeparttable}
\usepackage{float}
\usepackage{amsmath}
\usepackage{rotating}
\usepackage{hyperref}

\usepackage{relsize}
\usepackage{etoolbox}
\usepackage[most]{tcolorbox}
\usepackage{hyperref}
\usepackage{dcolumn}
\usepackage{bm}
\usepackage{wrapfig}
\usepackage{float}  
\usepackage[toc,page]{appendix}
\usepackage{gensymb}
\usepackage{epstopdf}
\usepackage{subfiles}
\usepackage{comment} 
\usepackage[autostyle=true]{csquotes} 
\usepackage{makecell}
\usepackage{pbox}
\usepackage{siunitx}
\usepackage[all]{hypcap} 
\usepackage[most]{tcolorbox}
\usepackage{hyperref}
\usepackage{xcolor}
\usepackage{nicefrac}
\usepackage{chemformula}
\usepackage{ulem}  
\definecolor{gainsboro}{rgb}{0.86, 0.86, 0.86}
\usepackage[backgroundcolor=gainsboro]{todonotes}
\hypersetup{
  colorlinks   = true, 
  urlcolor     = blue, 
  linkcolor    = blue, 
  citecolor   = blue 
}

\begin{document}

\title{}

\author{Sangeeta Rajpurohit}
\email{srajpurohit@lbl.gov}
\affiliation{Molecular Foundry, Lawrence Berkeley National Laboratory, USA}

\author{Jacopo Simoni}
\affiliation{Molecular Foundry, Lawrence Berkeley National Laboratory, USA}

\author{Liang Z. Tan}
\affiliation{Molecular Foundry, Lawrence Berkeley National Laboratory, USA}

\date{\today}

\title{Photo-induced phase-transitions in complex solids}

\begin{abstract}
Photo-induced phase-transitions (PIPTs) driven by highly cooperative interactions
 are of fundamental interest as they offer a
way to tune and control material properties on ultrafast timescales. Due to strong
correlations and interactions, complex quantum materials host several fascinating PIPTs
such as light-induced charge density waves and ferroelectricity and have become a desirable
setting for studying these PIPTs. A central issue in this field is the proper
understanding of the underlying mechanisms driving the PIPTs. As these PIPTs are highly nonlinear processes
and often involve multiple time and length scales, different theoretical approaches are often
needed to understand the underlying mechanisms. In this review, we present a brief
overview of PIPTs realized in complex materials, followed by a discussion of the available theoretical
methods with selected examples of recent progress in understanding of
the nonequilibrium pathways of PIPTs.  
\end{abstract}

\maketitle

\section{Introduction}
Over the past few decades, advances in ultrafast science have resulted in promising
routes to manipulate and control the properties of quantum materials on femtosecond timescales.
Photo-induced phase-transitions (PIPTs) in complex solids have emerged as one of the most
exciting fields in photonics and ultrafast science. \cite{Fausti2011,Giannetti2016,Keimer2017,Torre2021}.
These PIPTs on ultrafast timescales can be triggered by photo-induced nonthermal carrier
populations or charge fluctuations, or also by direct optical excitation of phonon modes. 
Complex quantum materials such as transition-metal oxides and layered van der Waals systems
belong to the class of materials with correlated electrons that interact with other degrees of
freedom. Recent ultrafast studies of PIPTs in complex quantum materials have provided a
new paradigm for controlling properties such as magnetism, metal-insulator transitions,
ferroelectricity, etc.

Here, we present a brief review of the recent progress in the field of ultrafast sciences,
particularly PIPTs in complex systems. We present selected
examples of experimental and theoretical studies of the photo-induced metal-insulator, magnetic,
structural phase-transitions that illustrate the phenomenology of these PIPTs and how they can
be understood in terms of their driving mechanisms and couplings between disparate degrees of freedom.
For a broad review of progress in photoinduced phenomena in quantum materials, we
refer to Torre et. al~\cite{Torre2021}, and we refer to  Koshihara et al. for a review experimental studies of PIPTs\cite{Koshihara2022}.
An overview of PIPTs in charge-transfer organic salt is provided by Onda et. al \cite{Onda2014}.\\
\begin{figure}[tp!]
     \begin{center}
     \includegraphics[width=0.95\linewidth]{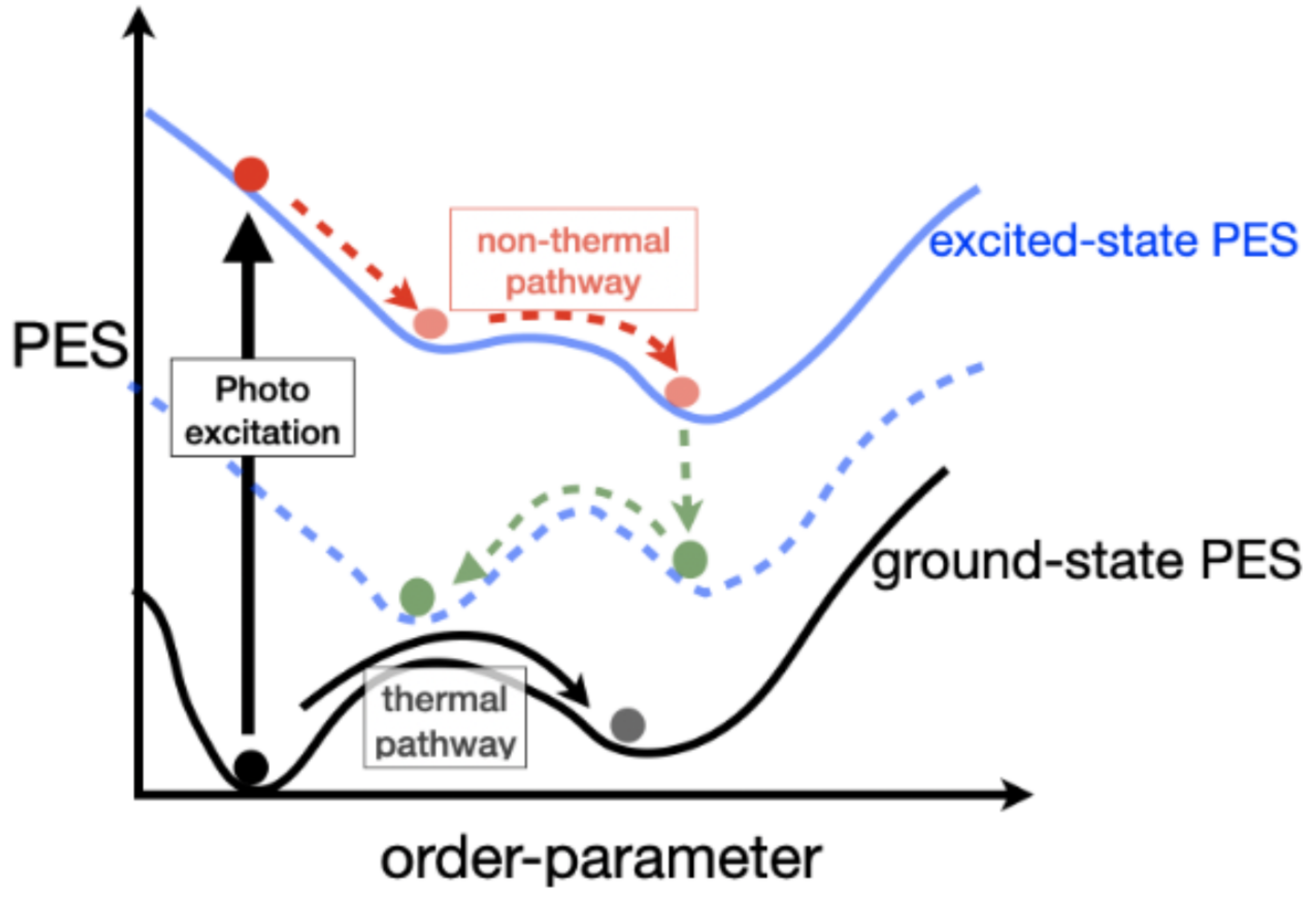}
     \caption{Illustration of changes in the 
     potential energy surface (PES) 
     during phase-transitions.}
     \label{fig:fig1}
     \end{center}
\end{figure}
\section{Materials hosting PIPTs}
Complex quantum materials exhibit rich equilibrium phase diagrams as a result of the
intimate coupling between electron, spin, and lattice degrees of freedom. The subtle
balance among different interactions in these systems drives fascinating ordering phenomena
such as magnetism, periodic lattice distortions, superconductivity, etc., resulting in
several intriguing broken symmetry and exotic states. PIPTs can be broadly classified
as those involving electronic, magnetic, or structural order parameters, or some
combination thereof. \\

Figure \ref{fig:fig1} illustrates the physics of PIPTs. The light illumination excites
electrons out of the ground state and drives the system into a new non-equilibrium state.
This new state can be a transient state or a metastable state with a long lifetime. The
excited system can also partially relax to a lower-energy configuration on the excited
state potential energy surface. Unlike quasi-equilibrium phase-transitions driven by changes
in temperature, pressure, and volume, PIPT are highly non-equilibrium in nature, and their
study requires a different set of experimental and theoretical techniques. New phases emerge
above threshold excitation fluences, at resonant excitation frequencies and at characteristic
time scales. Compared to thermal phase-transitions, PIPTs occur at ultrafast timescales, 
and host out-of-equilibrium "hidden" phases, which remain unattainable under equilibrium
conditions. Furthermore, the frequency- and polarization-sensitive coupling between the system
and light opens possibilities for ultrafast manipulation and storage of information. 

Recent ultrafast experimental studies have discovered phase-transition pathways
between these symmetry-broken states \cite{Fausti2011,Giannetti2016,Keimer2017,Torre2021}.
Moreover, selective modulation of electronic, lattice, and spin degrees with optical
excitations disentangles their individual contribution to the formation of these states. 
Pump-probe spectroscopy with femtosecond lasers measures transient changes in the system
induced by photoirradiation. Complementary experimental probes, such as time-resolved
diffraction and angle resolved photoemission spectroscopy (ARPES) are available to monitor electronic,
structural, and magnetic behavior during and after the PIPT \cite{Siders1999,Damascelli2003,Tinten2003,Fritz2007,Gedik2007,Sie2019}. \\

A prototypical example of an electronic PIPT is photo-induced charge density wave (CDW)
melting or transformation. CDWs are characterized by simultaneous periodic modification
of the electron density and lattice structure, driven by mechanisms such as fermi-surface
nesting, strong electron-phonon coupling, etc \cite{Rossnagel2011}. The
characteristic wavelengths of these CDW supperlattices are 2-6 times the lattice constants 
with the coherent length scales of several nanometers. \\

Transition metal layered materials, such as 1T-TaS$_2$ and RTe$_3$, are typical hosts of
PIPTs involving CDWs. The study of these materials in non-equilibrium states has led to a
more detailed understanding of the mechanisms behind CDW formation \cite{Stojchevska2014,Vaskivskyi2015,Hollander2015,Kogar2020,Zong2019}, see Table I. 
$\rm{TbTe_3}$ is a member of the $\rm{RTe_3}$ family known to host Fermi-surface nesting-driven
CDW formation \cite{Laverock2005}. The Fermi surface geometries in these systems
includes large parallel sections that can be spanned by a single nesting vector $q$ which induces
an electronic instability towards band opening by adding new periodicity of charge density.
The slight in-plane anisotropy in these systems makes the c-axis the preferred direction of the
CDW order. Using time-resolved ARPES measurements, Schmitt et al. \cite{Schmitt2008} showed
photo-induced closure of the band-gap at the Fermi level and hindered melting of CDW in TbTe$_3$. 
The photoexcitation changes the electronic screening instantaneously which triggers ions dynamics
towards a new potential minima in the presence of screening carriers. The longer timescales
of the involved collective excited atomic vibrations explains the observed delay in the CDW melting
and demonstrate the role of electron-phonon interaction in the origin of the CDW formation.
A recent experimental pump-probe study on LaTe$_3$ reported a PIPT in a symmetry-broken state where
following the optical excitation, the CDW along the c-axis weakens and, subsequently, a different
competing CDW along the a-axis appears \cite{Kogar2020}. The nearly identical timescales of relaxation
of this new CDW and re-establishment of the original CDW point to strong competition between the two
orders due to the presence of topological defects, as suggested by a previous study \cite{Zong2019},
which suppresses the recovery of the original long-range CDW. \\

\begin{figure}[tp!]
     \begin{center}
     \includegraphics[width=0.95\linewidth]{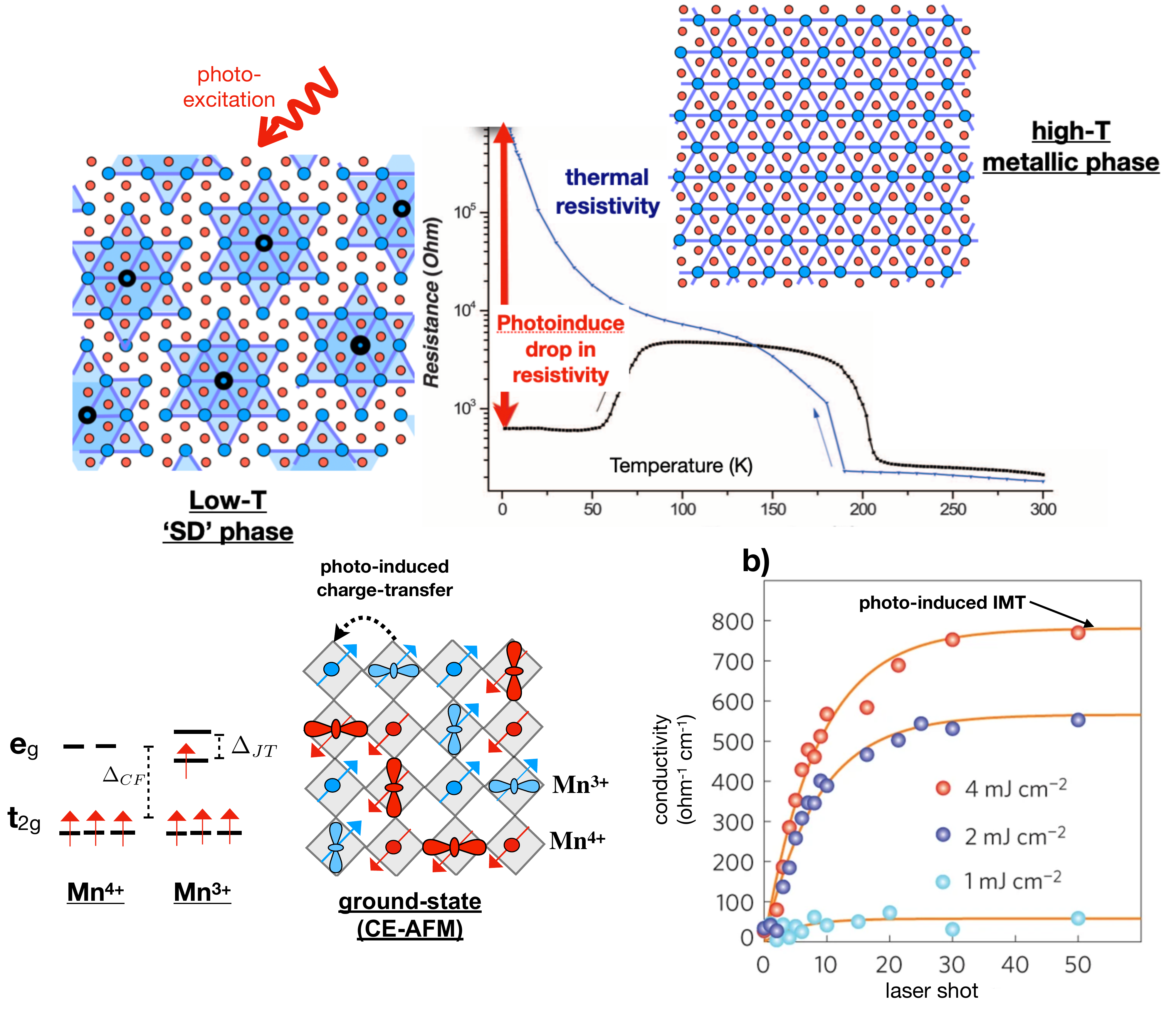}
     \caption{Top: Optical resistive switching in $\rm{1T{-}TaS_2}$. 
     The graph in center shows change in resistivity with
     temperature is shown in blue. The photoexcitation in
     low-temperature 'SD' phase reduces the resistivity,
     indicated in red arrow, by orders of magnitude. This figure has been adapted/reproduced from
      ref \cite{Stojchevska2014}  with permission from AAAS, copyright 2014.
     Bottom: The graph indicates step-like rise in conductivity in
     strain-engineered thin film of $\rm{La_{2/3}Ca_{1/3}MnO_3}$
     on photoexcitation at 80 K corresponding to 
     IMT. The CO, SO and
     OO pattern of $\rm{La_{2/3}Ca_{1/3}MnO_3}$ is
     shown in the left. Spin up and down Mn sites are
     indicated in blue and red.  This figure has been adapted/reproduced from ref \cite{Zhang2016} 
     with permission from Springer, copyright 2016.}
     \label{fig:fig2}
     \end{center}
\end{figure}

In other PIPTs, CDW formation is associated with insulator-metal transitions (IMT).
1T-TaS$_2$ exhibits a metallic phase (T-state) with an undistorted crystal structure at high
temperature and several commensurate and non-commensurate CDWs with decreasing temperature,
including the well-known low-temperature Star-of-David (SD) type pattern; see Figure \ref{fig:fig2}-top.
Previous theories suggesting that the low-temperature insulating CDW phase is Mott physics driven
have been challenged by recent studies that emphasize the role of orbital ordering intertwined with
CDW and their out-of-plane stacking \cite{Ritschel2015,Ritschel2018,Lee2019}. Ultrafast pump-probe
studies reveal photo-induced IMT to a new metastable hidden CDW state (HCDW), which has a metallic
character \cite{Stojchevska2014,Vaskivskyi2015,Hollander2015} The IMT may be linked
to the switching between these metastable states. A recent time-resolved XRD study identified 
the formation and breaking of interlayer dimer bonds between SD clusters \cite{Stahl2020} as a driving
mechanism of the phase-transition to the metallic hidden state. This is in agreement with
earlier experimental work suggesting ultrafast electronic timescales, involving charge and orbital
degrees of freedom, for phase-transitions \cite{Perfetti2006,Hellmann2010,Eichberger2010,Dean2011,Hellmann2012,Stojchevska2014,Tzong2015,Ligges2018}.\\

Instead of being accompanied by CDWs, photo-induced IMTs can also be accompanied by
structural phase-transitions. In VO$_2$, the photo-induced IMT is a transition from the dimerized
low-T monoclinic phase to the rutile phase of high-T \cite{Cavalleri2001,Cavalleri2004,Baum2007}.
However, the connection between these two phase-transitions and the underlying mechanism
remains unclear. Following early work  showing photoexcitation as another route to initiate IMT in
VO$_2$ \cite{Roach1971}, several ultrafast experimental attempts have been made to disentangle
the lattice and electronic contributions in the transition. These experiments suggest a slow
photo-induced IMT on picosecond time scales involving optical phonons, hinting at the lattice-assisted
Peierls-type mechanism \cite{Cavalleri2001,Cavalleri2004,Baum2007}. Recently, it has been 
suggested \cite{Morrison2014} that the change in crystal symmetry in the IMT  above the threshold fluence occurs on
the picosecond time scale, which is in agreement with earlier studies \cite{Cavalleri2001,Cavalleri2004,Baum2007}, 
this is due to a displacive transition in which all atoms collaboratively reshuffle themselves
in the displacive mode. However,  an analysis of x-ray scattering data, including the diffuse
continuum, \cite{Wall2018} suggests that the transition is a result of an order-disorder transition,
where the atoms move from the low- to the high-symmetry structure in a spatially incoherent manner. \\

The involvement of the spin degree of freedom further enriches the phenomenology of PIPTs. This is
exemplified in strongly correlated transition-metal oxides, such as manganites and nickelates.
While multiple-order parameters, linked to structural, electronic, and magnetic degrees of freedom,
are already present during thermal phase-transitions, the selective perturbation of these order
parameters through optical excitations  leads to new types of transformations
\cite{Rini2007,Li2013,Caviglia2013,Zhang2016, Esposito2018,Beyerlein2020,Stoica2020}.
In the Weyl semimetal system WTe$_2$, shear strain is coupled to ultrafast switching of topological
invariants~\cite{Sie2019}. The form of the PIPTs depends sensitively on the details
of the materials system. Ground state antiferromagnetic (AFM) insulating phase exhibiting 
long-range charge-order (CO) and orbital-order (OO) is transformed
to a long-lived metastable hidden metallic phase by resonantly exciting intersite $\rm{Mn^{3+}\rightarrow Mn^{4+}}$
transitions in a strain-engineered $\rm{La_{2/3}Ca_{1/3}MnO_3}$ \cite{Zhang2016},
see Figure \ref{fig:fig2}-bottom. The $\rm{Mn^{3+}\rightarrow Mn^{4+}}$ excitations-driven relaxation
of Jahn-Teller modes changes the lattice symmetry and affects the exchange-integral, which preserves
the itinerancy of Mn-d electrons. In comparison, AFM order is transformed to ferromagnetic (FM) order
above threshold excitation fluence in $\rm{Pr_{0.7}Ca_{0.3}MnO_3}$, attributed to fast non-equilibrium
spin canting on shorter timescales than Jahn-Teller and breathing phonon periods \cite{Li2013}.
 \\
\begin{figure}[tp!]
     \begin{center}
     \includegraphics[width=0.95\linewidth]{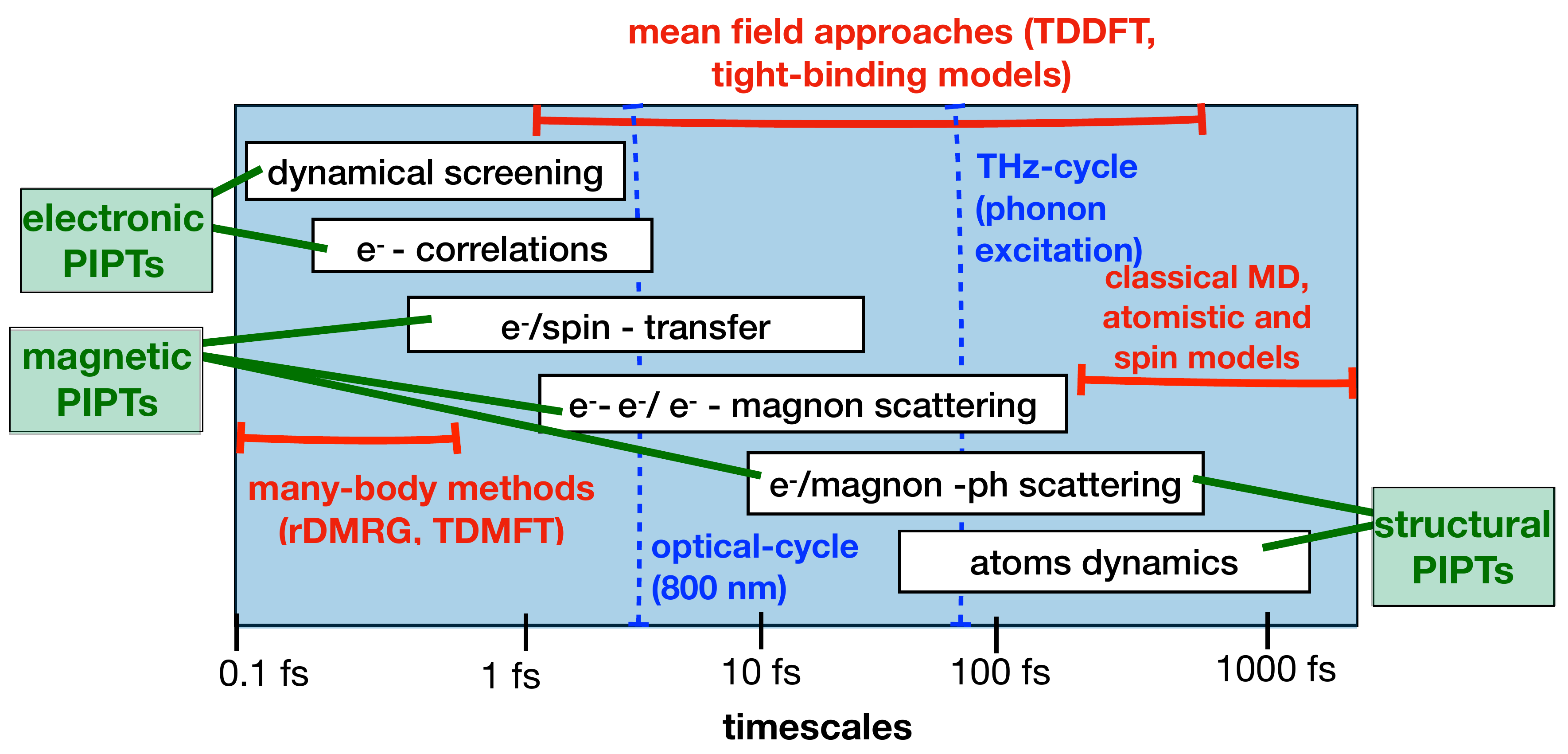}
     \caption{Timescales of elementary excitations and decay  processes in complex quantum materials.
     The blue dashed lines indicate the time-period of single-cycle of optical pulse and THz electric
     field which excites electrons and phonons, respectively. This figure has been adapted/reproduced from ref \cite{Petek1997} and \cite{Bovensiepen2012}
     with permissions from Elsevier and John Wiley and Sons, copyrights 1997 and 2012.} 
     \label{fig:fig3}
     \end{center}
\end{figure}

Even in laser-induced demagnetization processes which involve only the change of a
single magnetic order parameter, the mechanism can potentially involve many other
degrees of freedom. The processes of non-thermal laser-induced demagnetization can
be distinguished between photomagnetic where the photons excite the electrons to
states that have a direct influence on the magnetic properties; and optomagnetic
where the magnetization change is induced by a Raman type mechanism. \\
Following the first experimental observation of laser-driven change of the magnetic
state in ferromagnetic nickel \cite{Beaurepaire1996}, the ultrafast control of
the magnetic state has been achieved not only in the case of ferromagnetic and
ferrimagnetic metals, but also in the case of ferromagnetic semiconductors \cite{Wang2006}
and dielectrics \cite{Kirilyuk2006}.\\ 

Nonetheless, there is still uncertainty on the mechanism ultimately responsible for
the laser-induced change of magnetic state on such a short time scale and several
different driving forces have been proposed. These can be classified in at least
three main categories, a) Out of equilibrium modification of the electronic
and magnetic correlations \cite{Acharya2020,Mueller2013}; b) relativistic effects
and direct coupling between spins and laser \cite{Hinschberger2012}; c) coupling
to lattice or magnon degrees of freedom \cite{Carva2011,Carpene2008}. In general,
many of these effects could be at play at the same time and the observed demagnetization
is the result of their combined effect making the numerical modeling of the phenomenon
extremely complex to carry out.\\ 

Many experiments have confirmed that laser-induced demagnetization in ferromagnetic
transition metals takes place on a time scale of the order of approximately $500\,{\rm fs}$ or
less \cite{Kirilyuk2010,Acharya2020}. This has been demonstrated in the case of
${\rm CoPt_3}$ \cite{Guidoni2002}, thin ${\rm Fe}$ and ${\rm Ni}$ films \cite{Beaurepaire2004,Hilton2004}. 
In all these experiments the process was followed by radiation emission in the
Terahertz range. \\ 

Carrier induced ferromagnetism in semiconductors, that is caused by the magnetic
exchange interaction between localized spins and the itinerant charge carriers
(sp-d exchange interaction), provides an interesting alternative to the metallic
ferromagnets for magnetization control \cite{Ohno2000}. One of the first observations
of photo-induced demagnetization in semiconductors was reported in the case of GaMnAs \cite{Kojima2003},
but it occurred on a time scale of $1\,{\rm ns}$. More recently demagnetization on a
time scale of $1\,{\rm ps}$ was reported in the case of InMnAs by using combined Terahertz
and infrared excitations \cite{Mashkovic2020}. 
In the case of ferrimagnetic garnet films, instead, optical control of the magnetization
was shown to be possible on a time scale as short as few hundreds femtoseconds \cite{Kirilyuk2006}.\\

Besides inorganic solids, several organic molecular
complexes are also known to host PIPTs, for example, charge transfer (CT) molecular
complexes \cite{Iwai2002,Ikegami2007,Onda2008, Ishikawa2009,Kawakami2009,Uemura2010,Onda2014}.
The CT complexes belong to the class of strongly-correlated organic crystals which
consist of two types of $\pi$-conjugated molecules: an electron donor
and an electron acceptor. 
Increased electronic interactions due
to small overlap between $\pi$-orbitals of constituent donor and acceptor molecules
make these strongly-correlated materials. The PIPT was first discovered in CT organic
TTF-CA (tetrathiafulvalene-p-chloranil) \cite{Koshihara1990,Tanimura2004,Onda2014}, which
consists of chains of alternating electron-donor TTF and electron-acceptor CA molecules. 
The CT during photoexcitation in TTF-CA drives a
PIPT which involves switching from a neutral to an ionic state \cite{Iwai2002}.  (EDO–TTF)$_2$PF$_6$ is
another organic CT salt with strong el-el and el-ph coupling that exhibits a PIPT \cite{Fukazawa2012}. In
this system, EDO-TTF is a donor and PF$_6$ is an acceptor. (EDO–TTF)$_2$PF$_6$
has a low-temperature insulating phase with a (0, +1, +1, 0) charge-distribution
where `+1” indicates a hole on the EDO-TTF molecule and `O” means a neutral
EDO-TTF molecule. At room temperature, the system is metallic where EDO-TTF
carries an average charge of +0.5. This system exhibits a photoinduced hidden-phase with a
 (+1, 0, +1, 0) charge-distribution. This photoinduced hidden phase is
driven by strong el-el interactions and cannot be achieved in thermal equilibrium
\cite{Onda2008,Onda2014}. As another example, reflection-type femtosecond pump-probe
spectroscopy was used to detect photoinduced melting of the spin-Peierls phase in
the organic spin-Peierls alkali (M=K,Na)-tetracyanoquinodimethane (TCNQ) complexes \cite{Ikegami2007}.

\begin{table*}[!htb]
\label{tab:t2}
\centering
\begin{tabular}{|l l l l|}
\hline\hline
 \pbox{20cm}{\textbf{Material}\\ \textbf{(initial phase)}} & \pbox{20cm}{\textbf{Electronic/}\\\textbf{magnetic PIPT}} & \bf{Structural PIPT} & \bf{hidden-phases}  \\
\hline\hline
\makecell[l]{\textbf{VO$_2$}\\ (Monoclinic phase) } &
\makecell[l]{Insulator-to-Metal \\ 800-nm \\ $\sim$2-9 mJ/cm$^2$  \cite{Morrison2014}}  &
\makecell[l]{Monoclinic-to-Rutile \\ 800-nm \\  >9 mJ/cm$^2$  \cite{Morrison2014}}  &
\makecell[l]{Monoclinic-metallic \\ 800-nm  \\ $\sim$2-9 mJ/cm$^2$   \cite{Morrison2014}}  \\
\hline\hline
\makecell[l]{\textbf{1T-TaS$_2$} \\ (SD-phase)}  &
\makecell[l]{CDW-melting\\  1.50 eV (50-fs pulse) \\ 0.1 electrons /SD  \cite{Perfetti2006}} &
\makecell[l]{PLD-melting\\  790-nm (30-fs pulse) \\  0.5 mJ/cm$^2$   \cite{Petersen2011}} &
\makecell[l]{Metallic-CDW \\  800-nm (50-fs pulse) \\  $\sim$ 1 mJ/cm$^2$  \cite{Stojchevska2014}}  \\
\hline\hline
\makecell[l]{\textbf{RTe$_3$} (R=Tb,La) \\ (CDW along $c$-axis)}  &
\makecell[l]{CDW-melting \\  1.50 eV (50-fs pulse) \\  $\sim$ 2 mJ/cm$^2$   \cite{Schmitt2008} } &
\makecell[l]{ } &
\makecell[l]{new CDW along $a$-axis \\  800-nm/50 fs \\  $\sim$10-12 mJ/cm$^2$  \cite{Kogar2020}} \\
\hline\hline
\makecell[l]{\textbf{RR'MnO$_3$} \\  (AFM \& CO-OO)}   &
\makecell[l]{AFM-to-FM \\ 1.55-eV (50-fs pulse) \\  $\sim$5.8 mJ/cm$^2$  \cite{Li2013}}&
\makecell[l]{$Pbnm$-to-$P2_1m$ \\ 800-nm (55-fs pulse)\\ >4 mJ/cm$^2$  \cite{Beaud2014} } &
\makecell[l]{ferromagnetic-metallic\\  800-nm/(30-50 shots) \\ 2-5 mJ/cm$^2$  \cite{Zhang2016}} \\
\hline \hline
\makecell[l]{\textbf{(EDO–TTF)$_2$PF$_6$}\\ \textbf{(organic salt)} \\ (0,+1,+1,0) \\ charge-distribution}   &
\makecell[l]{insulator-to--metal; \\ (0,+1,+1,0)-to-\\(0.5,0.5,0.5,0.5) \\ charge-distribution \cite{Chollet2005}} &
\makecell[l]{ } &
\makecell[l]{ new CO with \\ (+1,0,+1,0)  charge-\\distribution \cite{Onda2008,Onda2014}} \\
\hline \hline
\end{tabular}
\caption{Examples of photoinduced electronic, structural, and magnetic phase transitions and
hidden phases in VO$_2$, 1T-TaS$_2$, RTe$_3$, RR'MnO$_3$ (R=rare earth metal and R'=divalent
alkaline earth metal) and (EDO–TTF)$_2$PF$_6$ discussed in the present study  \cite{Morrison2014,Perfetti2006,Stojchevska2014,Schmitt2008,Kogar2020,Li2013,Beaud2014,Zhang2016,Chollet2005,Onda2008,Onda2014}.}
\end{table*}  

\section{Theoretical descriptions of PIPTs}
Technological applications of PIPTs require predictive descriptions based on clear
understanding of the underlying pathways. These pathways involve a wide range of
time scales (Figure \ref{fig:fig3}), from femtosecond light absorption electronic
screening, to initial phonon dynamics on subpicosecond timescales, and to subsequent
relaxation that sometimes last up to the order of nanoseconds.  There is no single
theoretical approach which can describe the evolution of the excitations covering
all these time scales. In particular, in systems with strong correlations,
predictive descriptions of PIPTs become more challenging. Here, we give a brief
overview of the existing theoretical methods to study nonthermal dynamics,
particularly PIPTs that are mainly driven by quantum effects. Figure
\ref{fig:fig3} illustrates the timescales of the physical processes involved
in PIPTs and the appropriate theoretical methods used to study these processes. 
A wide range of explicit time-dependent theoretical methods, such as dynamical
mean-field theory (DMFT), density-matrix renormalization group DMRG, and
time-dependent density functional theory (TDDFT), are employed for out-of-equilibrium
treatment of ultrafast processes. \\ 

The t-DMFT and t-DMRG approaches take into account many-particle effects and have
been successful in describing PIPTs driven purely by strong
electronic correlations and screening effects, such as the IMT in Mott insulators
and excitonic insulators. The basic idea behind DMFT \cite{Georges1996} is to explicitly
consider correlations within a local region while
treating all non-local effects with dynamical mean-field to include quantum fluctuations
\cite{Georges1996}. The remaining local problem then becomes a so-called quantum-impurity
problem. The non-equilibrium extension of this approach, t-DMFT, has recently been
applied to the study of PIPTs \cite{Freericks2006,Eckstein2010,Eckstein2011}.  \\
\begin{figure}[tp!]
     \begin{center}
    \includegraphics[width=0.95\linewidth]{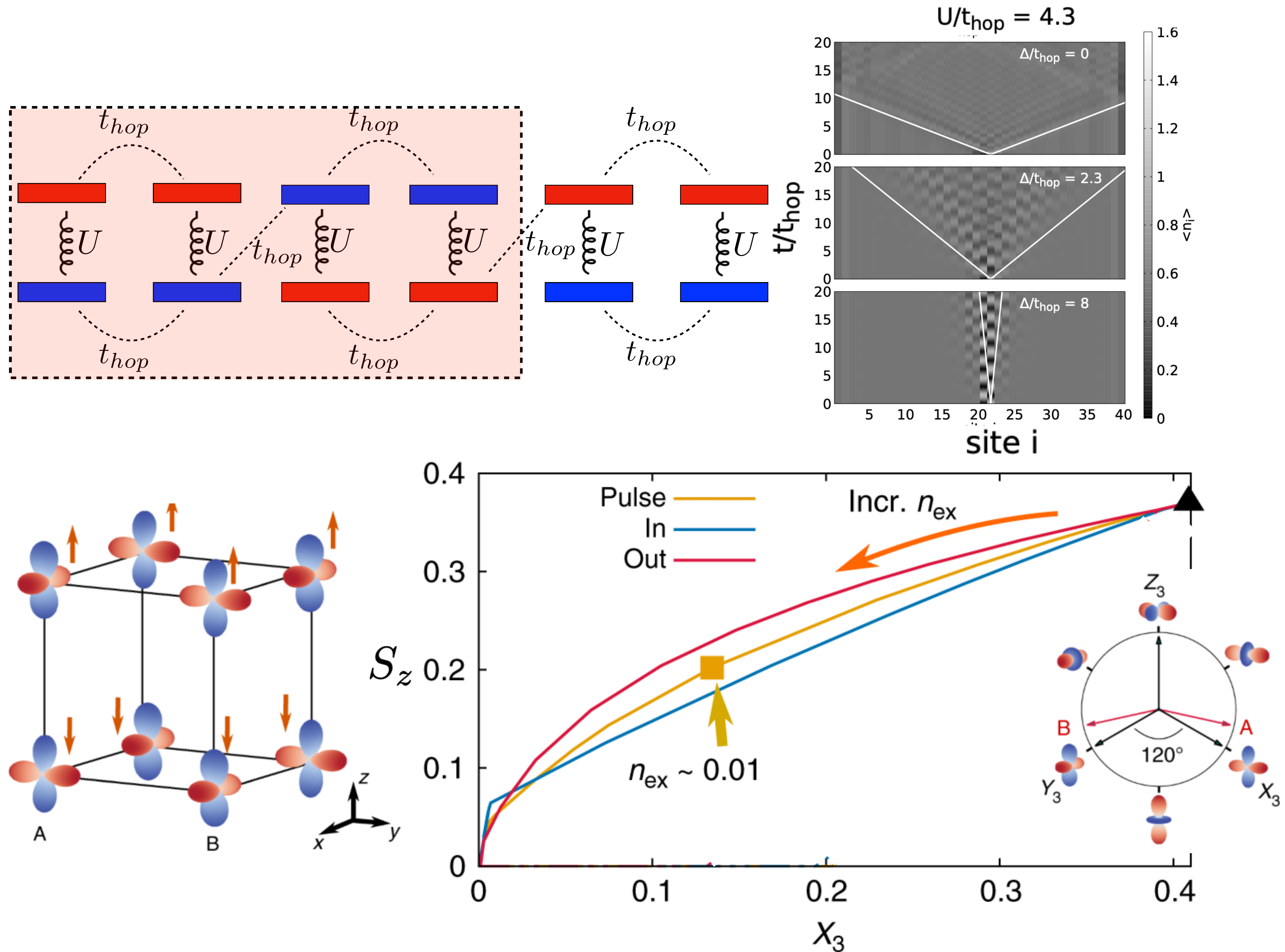}
    \caption{Top: Long-lived local excitations 
    in a dimerized chain induced by optical excitation, simulated with t-DMRG. Top-left: unit-cell with
    four-sites. Top-right: evolution of electron density after
    local excitation for different values of $\Delta/t_{hop}$,
    where $t_{hop}$ and $\Delta$ is hopping and onsite Hund’s splitting
    between opposite spin-orbitals, in a chain with 40 lattice sites.
    The local excitations are long lived at small $\Delta/t_{hop}$
    but spread with a "light-cone effect" at large $\Delta/t_{hop}$.
    This figure has been adapted/reproduced from ref  \cite{Kohler2018}
     with permission from APS, copyright 2018. Bottom: t-DMFT study showing photo-induced
    hidden-phase with new OO polarization in $\rm{KCuF_3}$.
    Bottom-left: SO and OO in the equilibrium state. Linear
    combinations of eg-states $|\theta\rangle$. Bottom-right: 
    Evolution of the total spin $S_z$ component and and OO,
    defined as the occupation difference $X_3=1/2(n_2-n_1)$ between 
    two eg-orbitals, in the long-time limit under three different non-equilibrium
    protocols, that is electric-field pulse (solid-pink), photo-doping
    electrons in (solid-yellow) and out (solid-blue). The inset shows the 
    $Z_1$-$Z_3$ plane in the orbital pseudospin space. The $X_3$, $Y_3$, $Z_3$
    directions and their corresponding orbitals are marked. This figure has been adapted/reproduced from ref \cite{Li2018}
     with permission from Springer, copyright 2018.} 
     \label{fig:fig4}
     \end{center}
\end{figure}
t-DMFT has revealed 
a photoinduced hidden phase with a new orbital-order
polarization in photoexcited KCuFe$_3$ (Figure \ref{fig:fig4}-bottom). Within
a 2-band Hubbard model, it as shown that the pathway to this
hidden state relies on a non-thermal partial melting
of the intertwined spin and orbital orders by
photoinduced charge excitations in the presence of
strong spin-orbital exchange interactions \cite{Li2018}. 
Another non-equilibrium DMFT study with the Hubbard-Holstein
model showed an ultrafast Mott-IM phase-transition
induced by optical excitation of coherent phonon modes.
The model revealed that nonlinear electron-phonon coupling is essential for this process \cite{Grandi2021}. 
While this approach has been successful for few-level models, the solution of the
non-equilibrium quantum impurity problem
is expected to become increasingly challenging as more states are added to the problem. 

The t-DMRG is a time-dependent extension of the original DMRG method to study non-equilibrum
processes \cite{Vidal2004,White2004,Schmitteckert2004,Daley2004}. Based on
Schmidt decomposition, where the system is divided into two blocks $A$ and $B$,
the equilibrium DMRG method \cite{White1992,White1993,Schollwock2005,Schollwock2011}
approximates the ground-state
wave function as
\begin{eqnarray}
|\psi\rangle=\sum_{i=1}^{\mathrm{dim}(H)}\alpha_i|\chi_i\rangle|\phi_i\rangle \approx \sum_{i=1}^m\alpha_i|\chi_i\rangle|\phi_i\rangle, m\ll\mathrm{dim}(H)
\end{eqnarray}
where, $m$ is very small. Here $|\chi_i\rangle$ and $|\phi_i\rangle$ are the
Schmidt bases of $A$ and $B$, respectively. The t-DMRG  extends the potential of the DMRG
method to study the non-equilibrum dynamics by solving the time-dependent Schr\"{o}dinger equation.
This approximation leads to a significant speed-up at the cost of truncating the wave
function, and allows treatment of systems which otherwise are not amenable to numerical or
analytical treatment. \\

t-DMRG has been applied to PIPTs where electronic correlation plays a crucial role
in the photoexcited state, such as photoexcitation of one-dimensional interacting
fermions in periodic magnetic and ionic microstructures\cite{Kohler2020}. With
magnetic microstructures, the spin-selective photo-excitation weakens the original
spin density pattern and a periodic charge modulation is induced. On the other hand,
with ionic potentials the periodic charge density pattern starts melting, and a
periodic modulation of the spin densities is induced. Similarly, long-lived bound
excitations have been shown to emerge in the  photoexcitated state of a one-dimensional
dimerized chain with Coulomb interactions \cite{Kohler2018}. Figure \ref{fig:fig4}-top
shows that the value of $\Delta/t_{hop}$, where $\Delta$ is the Hund's splitting,
controls the velocity of the spread of the excitation is seen to decrease, and
hence their lifetime. The study suggests that the underlying AFM magnetic microstructure
causes an increase in the relaxation times of excitations.

The application of t-DMRG to realistic systems with electron-phonon coupling has
been broadly successful \cite{Xie2019,Baiardi2019}, with examples such as the study
of carrier mobilities in organic semiconductors \cite{Li2020} or large-scale excitonic
dynamics in a system with electronic states coupled to hundreds of nuclear vibrations \cite{Borrelli2017}.
Although the t-DMRG approach is flexible and can be applied to a wide variety of electronic models
known for their quantum many-body effects, the approach is restricted mainly to one-dimensional systems.
For higher dimensions, the presence of intrinsic quantum entanglement
makes the truncation scheme unfavorable, restricting the accessible system size and accuracy.
\begin{figure}[tp!]
     \begin{center}
     \includegraphics[width=0.95\linewidth]{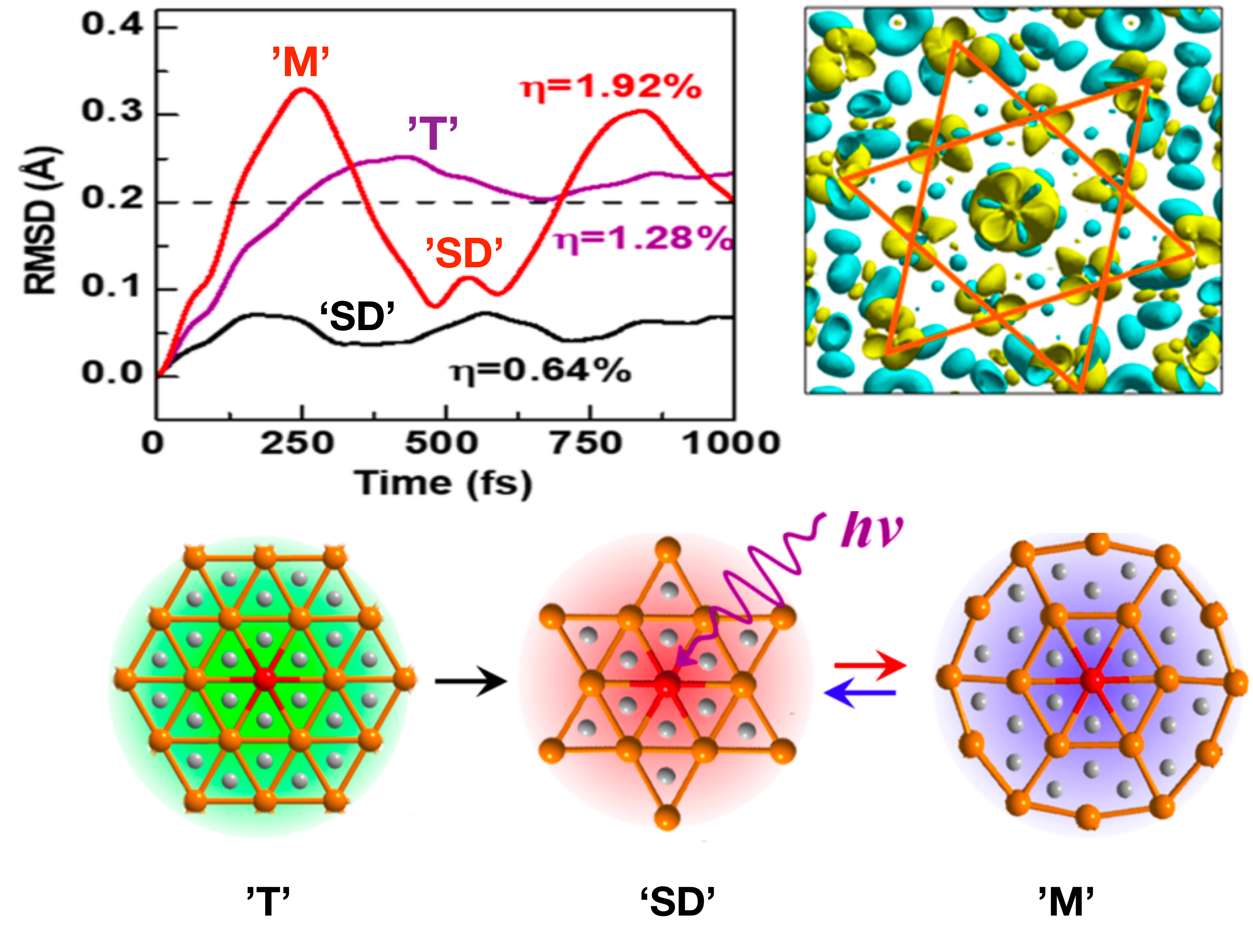}
     \caption{rt-TDDFT study of photo-induced melting of
     'Star-of-Davidson' (SD) pattern in $\rm{TaS_2}$.
     Top-left: Root mean square distance (RMSD)
     under three laser intensities, $\eta=0.64\%$ (black),
     $\eta=1.28\%$ (pink) and $\eta=1.92\%$ (red).
     The system retains original SD structure at
     $\eta=0.64\%$, but exhibits photo-induced metallic
     state 'T' at $\eta=1.28\%$. At high intensity $\eta=1.92\%$,
     the periodic oscillations suggests a new photo-induced transient
     metallic state 'M' with a new spatially-ordered atomic
     distortions. Top-right: light-induced
     charge-density redistribution at t=25 fs where yellow
     region shows increased density. Bottom:
     Nature of 'SD', 'T' and 'M' states.  
     This figure has been adapted/reproduced from ref  \cite{Zhang2019}
     with permission from ACS, copyright 2019. }
     \label{fig:fig5}
     \end{center}
\end{figure}

In complex materials, not only are electronic correlations strong,
but electron-phonon interactions can often play an important role as well. 
In the typical PIPT scenario involving electron-phonon coupling, photoexcitation first induces rapid changes in the
electronic structure, followed by modification of the
lattice structure to stabilize the new electronic order. 
As PIPTs in such systems involve large time and length scales
of phonon modes, studying them with highly accurate many-body
approaches discussed above is computationally unfeasible.
Instead, their description involves different electronic, magnetic, and lattice
pathways over long spatial and temporal scales. Several
mean-field methods incorporating electron-phonon interactions
have been developed for non-equilibrium phenomena in solids
at reduced computational cost with good accuracy. 

Real-time Time Dependent Density Functional Theory\cite{Runge1984,Gross1990} (rt-TDDFT)
is the time dependent extension of Density Functional Theory (DFT)\cite{Hohenberg1964,Kohn1965}.
This is a mean field approach where electrons individually interact
with the rest of the electronic system by means of an effective scalar potential,
the Kohn-Sham (KS) potential $v_{\rm KS}$. The single particle electronic wave
functions are evolved according to the following set of KS equations
\begin{equation} \label{eq:TDDFT}
    i\hbar\frac{\partial}{\partial t}\phi_i^{\rm KS}({\bf r},t)=\bigg[-\frac{\hbar^2}{2m}\nabla^2 + v_{\rm KS}[n]({\bf r},t)\bigg]\phi_i^{\rm KS}({\bf r},t)\,,
\end{equation}
\begin{equation} \label{eq:TDDFT2}
    v_{\rm KS}[n]({\bf r},t) = v_{\rm H}[n]({\bf r},t) + v_{\rm xc}[n]({\bf r},t) + v_{\rm ext}({\bf r},t)\,.
\end{equation}

The exact TDDFT xc-potential $v_{\rm xc}$ depends
on the initial interacting and Kohn-Sham states
(initial-state dependence) and the density at all previous times (memory) \cite{Casida2012}. 
However, most TDDFT studies use the adiabatic approximation, where one
assumes that the xc-potential reacts instantaneously and without memory to any
temporal change in the charge density. This adiabatic approximation to the
xc-potential fails to capture several photophysical and chemical processes 
\cite{Elliott2012,Helbig2011,Fuks2011}.Several attempts that goes beyond
the adiabatic approximation, including the time-dependent current density
functional theory (TDCDFT) \cite{Vignale1996,Vignale1997}, 
time-dependent deformation functional theory (TDDefFT) \cite{Tokatly2003,Ulrich2006} etc.,
have been made to include memory effects in the xc-potential. Recently, the
linear-response TDDFT nonadiabatic XC potential (XC kernel)
was calculated with DMFT for spin-independent and spin-dependent cases
\cite{Turkowski2017,Acharya2020}. In these studies, the kernels were derived
from the DMFT charge and spin susceptibilities, respectively.

The simplest rt-TDDFT approach ignores ion dynamics while evolving the electron
wavefunctions according to the quantum mechanical equations of motion. Such an approach
is suitable only for a time scale of the order of a hundred femtoseconds when the atomic
displacements can be approximated to be small. For this reason rt-TDDFT simulations have
been widely applied to study the ultrafast magnetization dynamics in transition metal
ferromagnets \cite{Krieger2015,Krieger2017,Dewhurst2021}, Heusler alloys\cite{Elliott2016}
and anti-ferromagnets\cite{Simoni2017}. In order to reproduce these phenomena the basic
rt-TDDFT equations (\ref{eq:TDDFT}) and (\ref{eq:TDDFT2}) are written in spinorial form
to account for spin non-collinearity\cite{Vignale2002}.

Ab-initio molecular dynamics (MD) can be divided in two main classes, namely
the adiabatic and the non adiabatic approaches. Both classes of methods
treat the electrons quantum mechanically, while the atoms obey the Newton's
laws of motion. 

The adiabatic methods are based on the so-called Born-Oppenheimer approximation\cite{Born1927}.
This approximation assumes that due to the difference in mass between the electrons
and the ions the electronic dynamics occurs on a much faster time scale compared to
the ionic one. At this level of approximation the electrons follow adiabatically the
atom dynamics, however effects like atomic induced electronic excitation
are neglected. For this reason the approach has limitations that become
particularly relevant in the description of out of equilibrium processes
typical of PIPTs. To this class belongs approaches like Born Oppenheimer Quantum MD
that have been employed to study non-thermal PIPTs in semiconducting systems like silicon and germanium\cite{Ji2013,Medvedev2013,Zijlstra2008,Medvedev2015,Silvestrelli1996}
and Car-Parrinello MD \cite{Car1985}.
The main limitation of the adiabatic approaches
comes from the assumption that the potential field on which the atoms move
can be approximated with the ground state electron energy surface. The
introduction of non adiabatic effects in the atomic motion has been hotly
debated for many decades\cite{Tully2012}. In general the atomic motion can
induce transitions between adiabatic states, in turn, electronic excitation
induces back-action on the atoms. These effects are particularly important in
the case of PIPTs when the electronic subsystem is highly excited by the
application of the laser field. The two most important non-adiabatic
methods are a) Ehrenfest quantum MD \cite{Li2005,Parandekar2006}
and b) Trajectory surface hopping method\cite{Tully1990}. 

Here we focus on Ehrenfest MD that is mostly used in the case
of solids. This approach is based on the solution of two sets of coupled
equations
\begin{eqnarray}
    i\hbar\frac{\partial}{\partial t}\psi_i({\bf r},t) & = & H({\bf r},\{{\bf R}_{\rm I}\},t)\psi_i({\bf r},t)\,,\\
    M_{\rm I}\ddot{\bf R}_{\rm I} & = & -\boldsymbol{\nabla}_{\rm I}E[\rho({\bf r},t)]\,.
\end{eqnarray}
Where $M_{\rm I}$, ${\bf R}_{\rm I}$ are ionic masses and positions,
$\rho({\bf r},t)$ is the electronic density, $E$ is the instantaneous energy surface
and $H({\bf r},t)$ is the electronic Hamiltonian where the electron-electron
interactions could be approximated as an exchange-correlation mean field potential. 
In this second case we have rt-TDDFT based Ehrenfest dynamics that has become
widespread in the study of PIPTs. 

rt-TDDFT combined with Ehrenfest dynamics has been employed to study the
ultrafast melting of silicon\cite{Lian2016}, photoexcitation induced charge
density waves\cite{Zhang2019}, PIPTs in transition-metal dichalcogenides\cite{Liu2020}
and semiconductors\cite{Bang2016}. rt-TDDFT has been used  \cite{Liu2020} to study the electronic and 
structural dynamics following photo-excitation in the low temperature IrTe$_2$ phase which exhibits
a periodic lattice distortion. Their study reveals that the microscopic force that
initiates the photo-induced structural dynamics and phase-transition in IrTe$_2$ 
originates from partial relaxation of the excited electrons of the anti-bonding
states of the Ir-Ir dimer in the conduction band by lowering their energy levels. 

A rt-TDDFT study of TaS$_2$ revealed melting of the CDW and 'SD' pattern in
photo-induced TaS$_2$ on femtosecond time scales, see Figure \ref{fig:fig5}.
The study further suggests that such photo-induced melting of the 'SD' pattern
cannot be explained by the hot electron model, as it is driven by collective
mode excitation due to intrinsic electron-nuclei coupled dynamics. A new
photo-induced transient metallic state was found with spatially ordered atomic
structures, different from 'SD' pattern. 

An important limitation of most non-adiabatic molecular dynamics
methods is the classical treatment of atoms and the inadequate description of quantum
coherence and decoherence effects \cite{Curchod2013}. For example, theoretical studies
of photoexcited state dynamics with full quantum-mechanical treatment of the electrons
and atoms highlight the importance of nonadiabatic transitions between vibronic states
in the photoinduced cooperative phenomena \cite{Ishida2008}.
Ehrenfest dynamics completely neglects quantum entanglement between the classical
and quantum degrees of freedom, while the independent classical-trajectories in the original
surface hopping method suffers from artificial overcoherence \cite{Tully1990}. To address
these issues, several methods have been proposed to add decoherence in the original
surface-hopping scheme \cite{Hack2001,Webster1991}.
\begin{figure}[tp!]
     \begin{center}
     \includegraphics[width=0.95\linewidth]{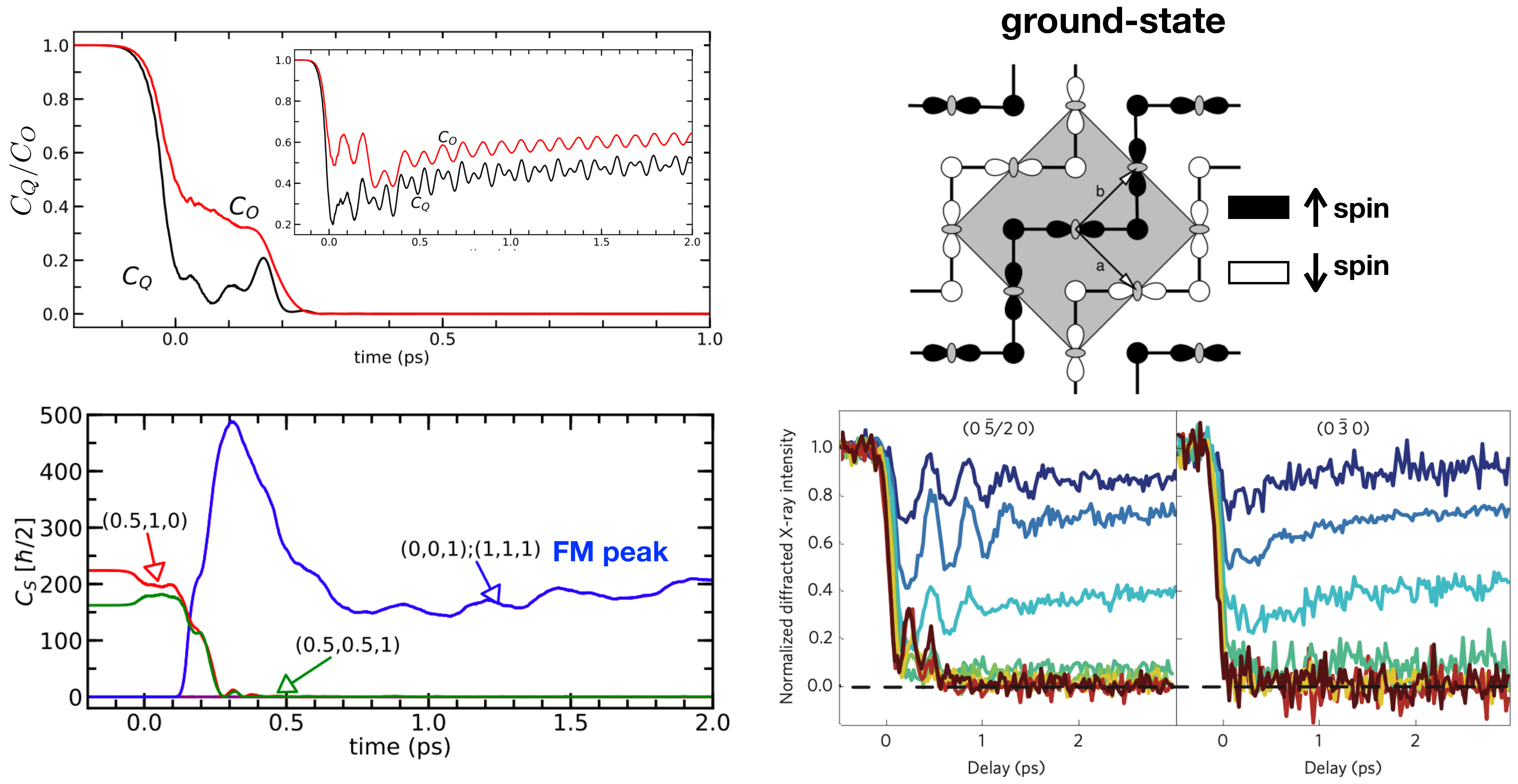}
     \caption{Tight-binding model study of 
     photo-induced FM-metallic phase in CO and OO antiferromagnetic
     $\rm{Pr_{1/2}Ca_{1/2}MnO_3}$. Right-top: CO, SO,
     and OO in equilibrium state. Left-top: Evolution of CO and OO peak as
     at high and low (inset) fluence value. For higher
     fluences, CO and OO melt. Left-bottom: Melting of the
     original SO (corresponding magnetic peaks are
     in green and red) and photo-induced FM order. Right-bottom:
     Experimentally study showing melting of CO (left) and OO (right)
     at high fluences. This figure has been adapted/reproduced from ref \cite{Beaud2014} and \cite{Rajpurohit2020_1}
     with permissions from Springer and APS, copyrights 2014 and 2020.}
     \label{fig:fig6}
     \end{center}
\end{figure}

Besides the ab-initio rt-TDDFT approach, the mean-field treatment of
model Hamiltonians have also been successful 
in understanding the ultrafast response that take place during PIPTs. 
One approach is to build an effective tight-binding model 
 with charge, spin, orbital and lattice
degrees of freedom explicitly included, and with all relevant interactions such 
as Coulomb interactions, electron-phonon coupling, 
Hund's coupling etc \cite{Rajpurohit2020_1,Rajpurohit2020_2} (Figure
\ref{fig:fig6}). Such models are required for PIPTs which involve multiple order parameters.
This study revealed melting of 
charge, orbital and lattice distortion in half-doped 
manganite $\rm{Pr_{1/2}Ca_{1/2}MnO_3}$ above a threshold fluence 
leading to a photoinduced metallic FM state. The emergence of FM 
order is attributed to optically induced spin-transfer (OISTR) 
between antiferromagnetically aligned ferromagnetic Mn chains. 
Such models aim to describe physics close to the Fermi level with
good accuracy and scale well for calculations of large
systems. In the field of ultrafast magnetization, model Hamiltonians with non-collinear
magnetic moments, spin-orbit coupling, and electron-phonon coupling have been
used to simulate photo-induced demagnetization \cite{Chen2019}.

As an example of an application that accesses longer length scales, ultrafast generation
of skyrmions was investigated within a 2-band electronic model with Rashba spin-orbit
coupling to a Heisenberg spin model. It was shown that the spin-orbit coupling was the
primary mechanism for the generation of these topological defects \cite{Bostrom2022}.

In some PIPTs, the coupling between electrons and spins, or between electrons and
phonons, only serves as an energy loss or decoherence mechanism instead of generating
multiple order parameters. For such PIPTs, it suffices to include these couplings as an
implicit dissipative effect. The femtosecond dynamics of magnetization in ferromagnetic
semiconductors\cite{Chovan2008} was studied using the Lindblad equation
of motion.
\begin{eqnarray}
i\partial_t\langle\rho\rangle=\langle[\rho,H(t)] \rangle
+ i\partial_t\langle\rho\rangle|_{relax}
\end{eqnarray}
in the mean-field approximation. The second term
in the above equation of motion introduces the
dephasing effect originating from hole spin-flip
interactions. A similar theoretical framework based on
non-linear density matrix equations of motion \cite{Li2013} was used to demonstrate that
transient ferromagnetic spin correlations can be driven entirely optically
in the femtosecond temporal regime. 

Purely atomistic models with only ionic degrees of freedom can be used to explain
PIPTs that involve only structural order parameters.  
For example, using a time-dependent atomistic model that 
embodies all the relevant phonon modes in manganites,
it was explained in Ref.~\cite{Beaud2014} that
the melting of the original charge- and orbital-order
together with Jahn-Teller distortions.
In this study, the coupling between the
experimentally determined time-dependent order
parameter $\eta(t)$, defined as the evolving electronic energy density,
and the phonon modes is used in the atomistic model 
to induce structural dynamics. The optical pulse changes
$\eta(t)$ that induced coherent oscillations of atomic motions. 

Finally, two (2TM) and three (3TM) temperature models are often
employed to capture the non-equilibrium and relaxation
dynamics after the PIPT for a few hundreds of picoseconds and beyond. The 2TM model\cite{Anisimov1974,Zahn2021}
is based on the assumption that the electrons and lattice
subsystems have well defined temperatures after the
first few hundreds femtoseconds of dynamics and that
only one parameter is required to model their energy
exchange (the electron-phonon relaxation rate). An alternative here is given by
extensions to the TTM based on the Boltzmann equation with the inclusion of
additional effects like the electron drifting\cite{Chen_2006}. In the
case of magnetic systems the 
Landau-Lifschitz-Gilbert (LLG) equation is usually employed
to evolve the spin degrees of freedom. The 3TM is
based on the same idea of the 2TM but extended to the
case of magnetic systems\cite{Roth2012}. The energy
is here allowed to flow between all three subsystems
electron, lattice, and spin.  A related approach involves the use of DFT with photoinduced
carrier populations modeled by Fermi-Dirac distributions, to approximate excited
state energy surfaces~\cite{Paillard2019}.

\section{Summary and Future outlook}
Ultrafast light-induced processes present a promising new frontier to create
exotic quantum phases that are unattainable under equilibrium conditions.
In this review, we provide a brief overview of recent experimental and
theoretical advances in ultrafast science, especially in PIPTs. The examples
discussed in the review illustrate the potential of PIPTs for tailoring material
properties on ultrafast timescales. With the development of X-ray free electron
lasers, future time-resolved inelastic resonant scattering (RIXS) techniques
\cite{Ament2011} of optically excited quantum systems with strong correlations 
and interactions have the potential to probe several low-energy quasiparticles such as
phonons \cite{Rossi2019}, magnons \cite{Lee2014,Zhou2013}, spinons \cite{Schlappa2012},
plasmons, excitons, charge excitations \cite{Hepting2018} etc., covering a wide range
of momentum and energy transfers. Recent experimental studies
with ultrafast electron diffraction (UED) \cite{Zhu2015,Konstantinova2018,Konstantinova2020,Li2022}
 have revealed transient electronic and lattice dynamics in materials
at an atomic scale. These studies show that electron pulses have become
a promising alternative to light and X-rays as an ultrafast probe \cite{Weathersby2015}
to investigate ultrafast dynamics at unprecedented resolution. 
The recently developed time-resolved photoemission electron microscopy (TR-PEEM)
is a powerful tool to visualize ultra-fast changes in materials with a temporal and spatial
resolution of a few femtoseconds and nanometers \cite{Kubo2005,Yu2008}.  TR-PEEM has
been applied to studying surface plasmon dynamics \cite{Kubo2007} 
and photo-excited carrier dynamics in semiconductors \cite{Fukumoto2014,Fukumoto2015}.

Experimental evidence of Floquet-Bloch bands in topological insulators suggests Floquet
engineering as a promising way to tailor properties \cite{Wang2013,McIver2020}. Recent 
observation of light-induced tuning of Moiré patterns in layered van der Waals
systems \cite{Vega2021} shows another example of tuning electronic and structural properties
on ultrafast timescales. Another developing area of research in ultrafast sciences is
photo-induced superconductivity motivated by the recent observation of transient
superconductivity in light-induced materials that are non-superconducting in the equilibrium
phase \cite{Fausti2011,Forst2015}. 

On the theoretical side, while there has been substantial progress in recent years,
new practical computational schemes for accessing large temporal and spatial scales
are highly desirable. Although TDDFT is formally an exact theory, the
adiabatic description of exchange and correlation has limitations, and better
approximations for exchange-correlation functionals need to be developed \cite{Sharma2011}.
As we have seen, the rt-TDDFT approach with Ehrenfest dynamics is sufficient
for describing some PIPTs. However, there are highly non-adiabatic photo-induced
processes where the ground- and excited state PES are markedly different, such as non-radiative
recombination and charge transfer. PIPTs which involve such behavior will likely require
beyond-Ehrenfest methods\cite{Horsfield2004, Kantorovich2018, Nijjar2019}. PIPTs which
span multiple time scales may require proper treatment of ultrafast excitation as well
as thermalization processes; the latter is another shortcoming of the Ehrenfest
method \cite{Rizzi2016}. Surface hopping methods \cite{Tully1990,Tapavicza2013,Curchod2018}
and their variants \cite{Shenvi2009,Roy2009_1,Roy2009_2} are a possible alternative
for modeling PIPTs with greater precision. The balance of accuracy and computational
efficacy is an important consideration in applying these algorithms to extended systems. 

Approximations to TDDFT, such as time-dependent tight-binding \cite{Todorov2001,Bonafe2020},
and other ab initio-based Hamiltonian models which include exchange and
correlation effects, scale well for larger systems and are potential methods for studying
correlated materials on longer time and length scales \cite{Rajpurohit2020_1,Rajpurohit2020_2, Bostrom2022}.
A point of consideration is in the proper parameterization of such models from ab
initio-based methods -- if only data from ground state calculations are used in the
parameterization, there is a possibility for errors to arise in the excited state PES.
Likewise, basis set choices that are appropriate for the ground state could potentially
be inadequate for the excited state if there is sufficient electronic delocalization
during the excitation process. Unlike the use of maximally localized Wannier
functions~\cite{Marzari2012} for ground state tight-binding, a canonical approach
to parameterizing excited state tight-binding models, particularly for PIPTs, is
not currently available.

While significant progress has been made in ultrafast science, there is still great scope
for improvements in experimental techniques, development of theoretical methods,
and exploration of new forms of PIPTs. This will deepen our understanding of non-equilibrium physics
and will offer a comprehensive understanding of the nonthermal pathways of
photo-induced phase-transitions in complex quantum materials.

\section{Acknowledgments}
S.R. was supported by the Computational Materials Sciences Program funded by the US
Department of Energy, Office of Science, Basic Energy Sciences, Materials Sciences
and Engineering Division. L.Z.T. and J.S. were supported by the Molecular Foundry, a DOE Office
of Science User Facility supported by the Office of Science of the U.S. Department of
Energy under Contract No. DE-AC02- 05CH11231.

\bibliography{ref}
\newpage
\end{document}